\documentclass[
 ,secnumarabic%
,amssymb,
nobibnotes,
nofootinbib,
aps, prl,
showpacs,showkeys]{revtex4}
\usepackage{docs}%
\usepackage{bm}%
\usepackage{graphicx}
\expandafter
\ifx
\csname package@font\endcsname\relax\else
\expandafter
\expandafter
\expandafter
\usepackage
\expandafter
\expandafter
\expandafter{\csname package@font\endcsname}%
\fi



\def\half{\frac{1}{2}}

\begin{document}

\title{Electromagnetic form factors of the nucleon:
new fit and analysis of uncertainties}
\author{W. M. Alberico$^1$}
\author{S. M. Bilenky$^{2,3}$}
\author{C. Giunti$^1$}
\author{K. M. Graczyk$^{1,4}$}
\email{graczyk@to.infn.it}
\affiliation{$^1$Dipartimento di Fisica Teorica, Universit\`a di Torino
and INFN, Sezione di Torino, I--10125 Torino, Italy}
\affiliation{$^2$BLTP, JINR, RU-141980 Dubna, Russia}
\affiliation{$^3$Physics-Department E15, Technische Universit\"{a}t M\"{u}nchen, D-85748 Garching, Germany}
\affiliation{$^4$Institute of Theoretical Physics, University of Wroc\l aw, pl. M. Borna 9, 50-204, Wroc\l aw, Poland}
\date{\today}%

\begin{abstract}
Electromagnetic form factors of proton and neutron, obtained from a new fit of data, are presented.  The proton form factors are obtained from a simultaneous fit to the ratio $\mu_p G_{Ep}/G_{Mp}$ determined from polarization transfer measurements and to $ep$ elastic cross section data. Phenomenological two-photon exchange corrections are taken into account. The present fit for proton was performed in the  kinematical region $Q^2\in (0,6)$~GeV$^2$. Both for protons and neutrons we use the latest available data. For all form factors the uncertainties  and correlations  of form factor parameters are investigated with the $\chi^2$ method.
\end{abstract}

\maketitle

\section{Introduction}
The nucleon electromagnetic form factors are fundamental quantities, of great theoretical and
experimental importance. The issue of their determination has been revisited in recent years, thanks to the results of several experiments at Bates, MAMI, JLab, which put under question previous analyses based on less precise data and urged the necessity for   a new parameterization and a new analysis  of the form factors themselves (for a review, see, for example Refs.~\cite{Gao:2003ag,Arrington:2006zm,Perdrisat:2006hj}).

A precise knowledge of the electromagnetic  form factors of the nucleon is  important for the determination of the axial nucleon form factor in charged current (CC) quasielastic neutrino-nucleon scattering \cite{Gran:2006jn} and strange form factors of the nucleon in
neutral current (NC) elastic neutrino-nucleon scattering. For example, NC vector form factors which characterize elastic NC scattering are given by the following expressions ~\cite{Alberico:1995bi}:
\begin{eqnarray}
&&G_E^{NC;p(n)} = \pm\half\left\{G_{Ep}-G_{En}\right\}
-2\sin^2\theta_W G_{Ep(n)} -\half{G_{Es}}
\nonumber\\
&&G_M^{NC;p(n)} = \pm\half\left\{G_{Mp}-G_{Mn}\right\}
-2\sin^2\theta_W G_{Mp(n)} -\half{G_{Ms}}\,.
\nonumber
\end{eqnarray}
In the above the dominant terms are the electric ($G_E$) and magnetic ($G_M$) form factors of the nucleon.
Their precise knowledge is essential in order to  determine the small strange form factors of the nucleon
$G_{Es}$, $G_{Ms}$. Hence, it is obvious that not only a good knowledge of the electromagnetic form factors
is required, but also the present level of their uncertainty.

In this paper we  performed   new fits of the nucleon electromagnetic form factors.
The proton  ones  are extracted from: i) elastic $ep $ cross section data, ii) polarization data, providing the $\mu_p G_E/G_M$ ratio ($\mu_p$ being the magnetic moment of the proton).
The neutron form factors are extracted from electron-nucleus (typically deuterium and $^3$He) scattering processes. The latest experimental data are used.

The proton form factors determined from the measurements of  polarization transfer in
elastic electron-proton scattering
(first appearing between '99 and '02) were in a  significant disagreement with respect to the ones obtained from elastic $ep$ scattering data via the customary Rosenbluth separation.
The main suggestion to solve  this inconsistency was to account for
two photon exchange (TPE) diagrams \cite{Guichon:2003qm,Blunden:2003sp,Chen:2004tw,Afanasev:2005mp}, which should affect the cross section to a greater extent than the polarization data.

This disagreement  became even more evident after the new JLab data on $ep$ scattering cross sections \cite{Qattan:2004ht} appeared. Hence, as already pointed out by several authors (see e.g. \cite{Arrington:2007ux}), a reliable global fit must include the TPE correction; we will explicitly show the effect of TPE on the goodness of the fit (GoF). For a recent review devoted to TPE correction see Ref. \cite{Carlson:2007sp}.

There exist several parameterizations of the nucleon form factors which have been considered in the literature
~\cite{Arrington:2007ux,Bosted:1994tm,Brash:2001qq,Budd:2003wb,Arrington:2003df,Arrington:2003ck,Kelly:2004hm,Lomon2001,Lomon:2006xb,Arrington:2006hm,Bodek:2007ym}. Among these, the older ones  have a purely empirical $Q^2$ dependence~\cite{Bosted:1994tm,Brash:2001qq,Budd:2003wb,Arrington:2003df}:
$$
G_{Ep}(Q^2), G_{Mp}(Q^2)/\mu_p \sim \frac{ 1}{\displaystyle 1 + \sum_{i=1}^N c_i Q^i}, \quad Q = \sqrt{Q^2}.
$$
The specific form of the parameterization may depend on the $Q^2$ region. For instance, in Ref.~\cite{Arrington:2006hm} the low-$Q^2$ data were analyzed with form factors given by a continued fraction parameterization:
$$
G_{Ep}(Q^2), G_{Mp}(Q^2)/\mu_p \sim \frac{1}{\displaystyle 1 + \frac{b_1  Q^2}{\displaystyle 1 + \frac{b_2 Q^2}{1+...}}}.
$$
The newest empirical form factors are constrained to have a proper physical behavior  at low-$Q^2$ as well as at high-$Q^2$. One of the examples is the Kelly's parameterization~\cite{Kelly:2004hm}, which will be employed in our analysis (see next Section). The form factors depend on powers of the invariant $Q^2$, and for large $Q^2$ the form factors behave like $1/Q^4$. In Ref.~\cite{Bodek:2007ym} Kelly's parameterization is additionally constrained
to satisfy duality hypothesis and the low-$Q^2$ behavior is described as in  Ref.~\cite{Arrington:2006hm}.

The electric neutron form factor, usually, is separately treated and
described with a smaller number of parameters~\cite{Galster:1971kv}
(see also \cite{Krutov:2002tp}).

It is also worth mentioning those parameterizations obtained on the basis of the vector meson dominance model.
In particular the parameterization proposed by  Lomon~\cite{Lomon:2006xb}
seems especially suited to successfully describe the neutron form factor data.

In the present paper we aim to provide  reliable fits of both proton and neutron e.m. form factors, by employing a relatively small number of parameters; moreover one of the major merits of this work is the analysis of errors on the parameters of the fit, which allows one to estimate the present uncertainty on our knowledge of the electromagnetic form factors.

The paper is organized as follows: in Section 2 we consider the proton form factors, by analyzing both the polarization data (Section \ref{polarization}) and the cross section data (Section \ref{cross section}). Section 3 is devoted to the neutron form factors, separately considering the electric and the magnetic form factors. Finally Section 4 presents a discussion of our results in comparison with previous analyses and the conclusions.

\section{Proton form factors}

In Section~\ref{polarization} we consider the recent polarization transfer and asymmetry measurements data which give an information on the
ratio of the electric and magnetic proton form factors.
Then, in Section~\ref{cross section},
we present the results of the combined fit of the  polarization and cross section data.

\subsection{Fit of polarization data}
\label{polarization}

\begin{figure}[t]
\centering{
\includegraphics[width=0.6\textwidth]{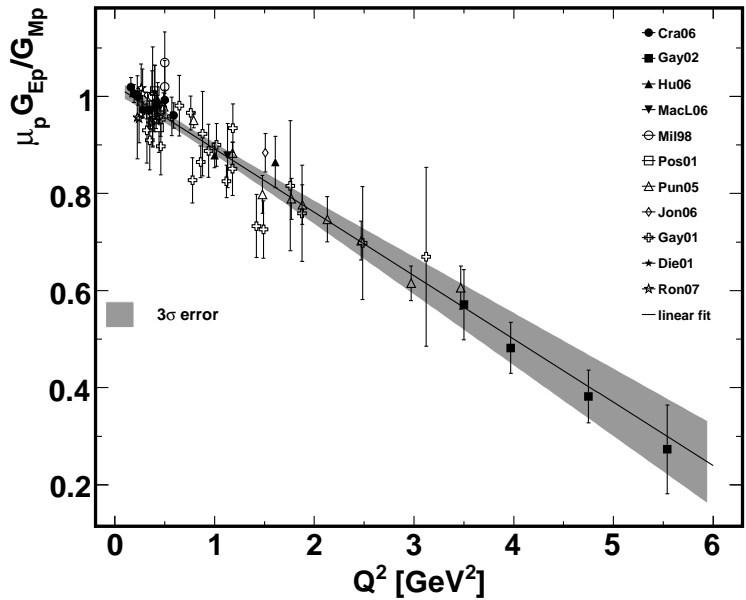}
\caption{ Linear fit (\ref{ratio-fit}) of the recoil polarization
and asymmetry measurements of the ratio $\mu_p G_{Ep}/G_{Mp}$. The
shadowed area denotes the $3\sigma$ C.L. region of the fit.
\label{fig_ratio_linear} }}
\end{figure}

In this section we consider the direct determination of the ratio of the electric and magnetic
proton form factors
\begin{equation}
\mathcal{R}(Q^2)
\equiv
\mu_p \, \frac{G_{Ep}(Q^2)}{G_{Mp}(Q^2)}
\,,
\label{ratio1}
\end{equation}
which has been obtained with the measurement of
the polarization of the recoil proton and with asymmetry measurements.
Here $ Q^2 \equiv - q^2 $, $q$ being  the four-momentum transfer.
In the one-photon approximation $q$ is the four-momentum of the virtual photon.

The recoil polarization technique
(see Ref.~\cite{Perdrisat:2006hj})
has been employed in several $ep$ experiments
for a direct measurement of the ratio $\mathcal{R}(Q^2)$. In the laboratory
frame it is given by:
\begin{equation}
\mathcal{R}(Q^2)
=
- \mu_p \, \frac{P_t}{P_l} \, \frac{E + E'}{2M} \, \tan\!\left(\frac{\theta}{2}\right)
\,,
\label{ratio2}
\end{equation}
where $P_l$ and $P_t$ are the longitudinal and transverse components of the
polarization of the recoil proton, $M$ is the proton mass,
$E$ and $E'$ are the initial and final electron energies,
and $\theta$ is the electron scattering angle. The latter is related to $Q^2$
according to:
\begin{equation}
Q^2 = 4 E E' \sin^2\left( \frac{\theta}{2} \right)
\,.
\label{theta-lab}
\end{equation}

The ratio $\mathcal{R}(Q^2)$ has been also determined from the measurement of
the asymmetry in elastic $ep$ scattering  with both polarized  beam and target:
we have (see Ref.~\cite{Perdrisat:2006hj})
\begin{eqnarray}
 \label{asymmetry}
\frac{ \sigma_{+} - \sigma_{-} }{ \sigma_{+} + \sigma_{-} } &=&
 - 2 \mu_p \sqrt{\tau(1+\tau)} \tan\!\left(\frac{\theta}{2}\right)
\frac
{ \mathcal{R} \sin\theta^{*} \cos\phi^{*} + \mu_p \sqrt{\tau[1+(1+\tau)\tan^2\left(\frac{\theta}{2}\right)]} \cos\theta^{*}}
{ \mathcal{R}^2 + \mu_p \tau / \epsilon }\,,
\end{eqnarray}
where $\sigma_{+}$ and $\sigma_{-}$
are the cross sections for positive and negative electron helicities, respectively,
$\theta^{*}$ and $\phi^{*}$
are the polar and azimuthal angles of the target polarization
relative to the three-momentum transfer vector $\vec{q}$ and the scattering plane
(in the laboratory frame),
\begin{equation}
\tau \equiv \frac{Q^2}{4 M^2}
\,,
\label{tau}
\end{equation}
and
\begin{equation}
\epsilon \equiv \left[1+2(1+\tau)\tan^2\!\left(\frac{\theta}{2}\right)\right]^{-1}
\label{epsilon}
\end{equation}
is the virtual photon polarization.

We consider the recoil polarization and asymmetry data published in
Refs.~\cite{Milbrath:1997de,Jones:1999rz,Dieterich:2000mu,Pospischil:2001pp,Gayou:2001qt,Gayou:2001qd,Punjabi:2005wq,Crawford:2006rz,Hu:2006fy,Jones:2006kf,MacLachlan:2006vw,Ron:2007vr}. These data are plotted in Fig.~\ref{fig_ratio_linear}, together with their error bars, which
include the statistical and systematic uncertainties added in
quadrature. The $Q^2$ range of the data goes from 0.15 to 5.6
GeV$^2$. As one can see from Fig.~\ref{fig_ratio_linear}, all data
are well described by a linear function in $Q^2$:
\begin{equation}
\label{R_linear}
\mathcal{R}(Q^2) = c_0 + c_1 Q^2.
\end{equation}
We fitted the data points with this linear function, by minimizing the least-squares
function
\begin{equation}
\chi^2_{\text{rat}} = \sum_{j=1}^{N_{\text{rat}}}
\frac{ ( \mathcal{R}(Q^2_j) - \mathcal{R}_j^{\text{exp}} )^2 }{ (\Delta\mathcal{R}_j^{\text{exp}})^2 }.
\label{chi2rat}
\end{equation}
where $ N_{\text{rat}} = 65 $ is the total number of recoil polarization and asymmetry data points
and $\mathcal{R}_j^{\text{exp}}$ is the value of the ratio at the
squared-momentum transfer $Q^2_j$, with corresponding uncertainty $\Delta\mathcal{R}_j^{\text{exp}}$.

We found the following best fit values of the parameters:
\begin{equation}
\label{ratio-fit}
c_0 = 1.022 \pm 0.005
\,,
\qquad
c_1 = - 0.130 \pm 0.005
\,,
\end{equation}
with 1$\sigma$ uncertainties computed from the covariance matrix\footnote{Detailed numbers of the covariance matrices relative to these and to the following fit parameters can be found in Ref.~\cite{website_results}.}
(they are given by the square-roots of the diagonal elements of the covariance matrix).

The corresponding minimum $\chi^2$ being:
\begin{equation}
(\chi^2_{\text{rat}})_{\text{min}} / \text{NDF} = 58.89 / 63
\,,
\label{chi2min-rat}
\end{equation}
where $ \text{NDF} = N_{\text{rat}} $ is the number of degrees of freedom;
the goodness of the fit (see Ref.~\cite{Amsler:2008zz}) is 62\%.

\begin{figure}[t!]
\centering{
\includegraphics[width=0.48\textwidth]{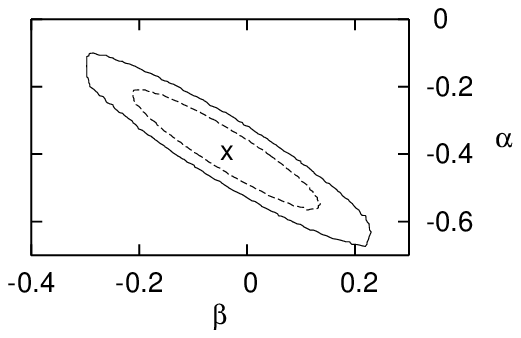}
\includegraphics[width=0.48\textwidth]{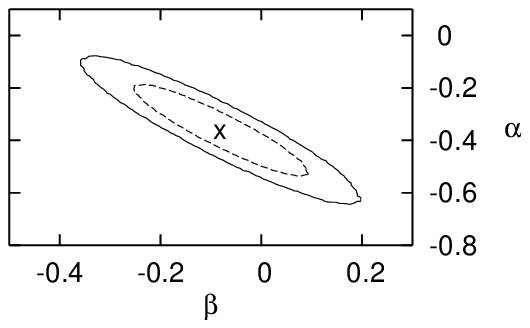}
\caption{ Projection on the $\alpha-\beta$ parameter space of the
contours delimiting the allowed regions in the 2-dimensional space
of TPE correction parameters, with $1\sigma$ (dashed lines) and
$2\sigma$ (solid lines) C.L. The contours are computed for fit~I
(left figure) and fit~II right figure. The crosses indicate the
projections of the best-fit point, Eq.~(\ref{par-glo-1-tpe}) for
fit~I and Eq.~(\ref{par-glo-2-tpe}) for fit~II. }
\label{fig_contour_tpe_r}}
\end{figure}

The solid line in Fig.~\ref{fig_ratio_linear} corresponds to the
best-fit values of the parameters in Eq.~(\ref{ratio-fit}), while
the shadowed area denotes $3\sigma$ C.L. region of the fit. One can
see that the linear fit has small uncertainties, especially for $
Q^2 \lesssim 3 \, \text{GeV}^2 $, where there are many data points.

The best-fit values of the parameters $c_0$ and $c_1$ in Eq.~(\ref{ratio-fit}) are
close to those obtained by Arrington in Ref.~\cite{Arrington:2003df},
$ c_0^{\text{bf}} = 1.0324 $ and $ c_1^{\text{bf}} = - 0.135 $. Let us notice that in Ref.~\cite{Arrington:2003df}  it was assumed that for $Q^2<0.24$~GeV$^2$ the form factor ratio is equal to one.


\subsection{Fit of polarization and cross section data}
\label{cross section}
\begin{figure}[t!]
\centering{
\includegraphics[width=0.8\textwidth]{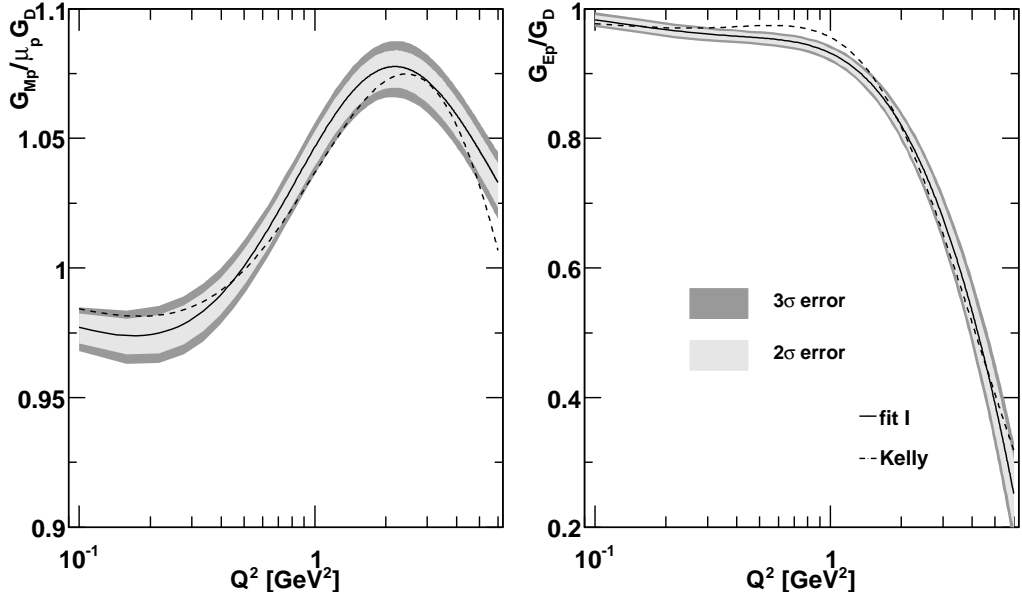}
\caption{ The fits  of $G_{Mp}/\mu_p G_{D}$ and $G_{Ep}/G_{D}$ form
factors. The solid lines denote fit~I. The shadowed areas represent
the $2\sigma$ (bright) and $3\sigma$ C.L. (dark) allowed regions.
Kelly's fit \cite{Kelly:2004hm} is shown by the dotted line.
\label{gmp_and_gep_with_ratio} }}
\end{figure}

The values of the proton form factors have been extracted from the data of
many elastic $ep$ scattering experiments using the Rosenbluth method.
In the one-photon approximation, the differential cross section in
the laboratory frame for unpolarized $ep$ elastic scattering reads
(in the same notation used in the previous subsection):
%
\begin{equation}
\label{cross_section_1PE}
\sigma(E,Q^2)
\equiv
\frac{d \sigma_{1\gamma}}{d\cos \theta}
=
\sigma_{\text{M}}(E,Q^2)
\left( G_{Ep}^{2}  + \frac{\tau}{\epsilon} \, G_{Mp}^{2}\right) \left( \frac{1}{1+\tau} \right)
\,,
\end{equation}
$\sigma_{\text{M}}$ being the Mott's differential cross section
\begin{equation}
\sigma_{\text{M}}(E,Q^2)
\equiv
\left( \frac{d \sigma}{d \cos\theta} \right)_{\text{M}}
=
\frac{ \pi \alpha^2 E' \cos^2(\theta/2) }{ 2 E^3 \sin^4(\theta/2) }
\,.
\end{equation}
The Rosenbluth separation is then obtained by considering the reduced differential cross section
\begin{equation}
\sigma_{\text{R}}(E,Q^2)
\equiv
\epsilon
\left( 1 + \tau \right)
\frac{ \sigma(E,Q^2) }{ \sigma_{\text{M}}(E,Q^2) }
=
\tau \, G_{Mp}^2(Q^2) + \epsilon \, G_{Ep}^2(Q^2)
\,.
\label{Rosenbluth}
\end{equation}

A linear fit  of the reduced differential cross section at fixed
$Q^2$ and different values of $\epsilon$ gives the value of $\tau G_{Mp}^2(Q^2)$ from the
intercept ($\epsilon=0$) and the value of $G_{Ep}^2(Q^2)$ from the slope.
Notice, however, that the measurement of $G_{Ep}^2(Q^2)$ with the Rosenbluth method
has large uncertainties, because the contribution of $G_{Ep}^2(Q^2)$ to the
reduced differential cross section in Eq.~(\ref{Rosenbluth}) is suppressed for
large values of $Q^2$ ($\tau\gtrsim\epsilon$) while for small values of $Q^2$
we have $ G_{Ep}^2 \simeq G_{Mp}^2 / \mu_{p} \simeq G_{Mp}^2 / 7.8 $.

In our analysis, in the first fit, later called \emph{fit~I}, we
assume that $G_{Ep}$ is related to $G_{Mp}$ by the linear relation of
Eq.~(\ref{R_linear}), which is favored by the direct measurement of
$\mathcal{R}(Q^2)$ in polarization experiments, as discussed in Section~\ref{polarization}.

For the proton magnetic form factors we adopt the parameterization proposed by
Kelly \cite{Kelly:2004hm}:
\begin{equation}
\label{Kelly_Form_Factor}
\frac{ G_{Mp}(Q^2) }{ \mu_p }
=
\frac{\displaystyle 1 + \sum_{k=1}^{n} a_{p,k}^{M} \tau^k }
{\displaystyle 1 + \sum_{k=1}^{n+2} b_{p,k}^{M}\tau^k }
\,,
\end{equation}
which guarantees the  asymptotic behavior $ G_{Mp}(Q^2) \propto
Q^{-4} $ at high $Q^2$ \cite{Brodsky:1973kr}. We shall employ the
parameterization of Eq.~(\ref{Kelly_Form_Factor}) with 4 parameters
($n=1$):
\begin{equation}
\label{Kelly_Form_Factor_discussion}
\frac{ G_{Mp}(Q^2) }{ \mu_p }
=
\frac{\displaystyle 1 +  a_{p,1}^{M}\tau }
{\displaystyle 1 + b_{p,1}^{M}\tau + b_{p,2}^{M}\tau^2 + b_{p,3}^{M}\tau^3}
\,.
\end{equation}
We will see that this choice turns out to be quite
satisfactory for the description of the data.
Moreover a relatively small number of parameters
allows a better control of the errors.

We have also performed a fit with both the magnetic and electric proton form factors
parameterized by the expression (\ref{Kelly_Form_Factor_discussion}).
This fit will be  called \emph{fit~II} in the following.
In this case the electric form factor reads:
\begin{equation}
\label{Kelly_GE_Form_Factor_discussion}
G_{Ep}(Q^2) =
\frac{\displaystyle 1 +  a_{p,1}^{E}\tau }{\displaystyle 1 + b_{p,1}^{E}\tau + b_{p,2}^{E}\tau^2 + b_{p,3}^{E}\tau^3}
\,.
\end{equation}

In our analysis we consider similar sets of cross section data as the ones employed by
Arrington in Ref.~\cite{Arrington:2003ck}, namely the data from Ref.~\cite{Janssens:1965kd}-\cite{niculescu_PHD}.
Some of the data were taken from the JLab data base~\cite{JLab_data}; we include also data from Ref.~\cite{Arnold:1986nq}.  Additionally, we considered the latest data of JLab experiment~\cite{Qattan:2004ht} in which the cross section was measured with the smallest errors, up to-date.

We fitted the $ep$ cross section data by minimizing the least-squares function
\begin{equation}
\label{chi2_cross_only}
\chi^2_{\text{cs}}
=
\sum_{i=1}^{M_{\text{cs}}}
\left\{
\sum_{j=1}^{N^{\text{cs}}_i}
\frac{ \left[ n_i\sigma_{i,j}^{\text{exp}} - \sigma(E_{i,j},Q_{i,j}^2) \right]^2 }{ (\Delta\sigma_{i,j}^{\text{exp}})^2 }
+
\frac{(1-n_i)^2}{(\Delta n_i)^2}
\right\}
\,,
\end{equation}
where $M_{\text{cs}}=24$ or $M_{\text{cs}}=28$ \footnote{For $Q^2\in(0.1,6)$ we have 24 independent data sets while for $Q^2\in(0,6)$ we have 28 independent data sets.} are the numbers of data sets,
$N^{\text{cs}}_i$ is the number of points in the $i$th data set,
$n_i$ and $\Delta n_i$ are the corresponding overall normalization and uncertainty,
$\sigma_{i,j}^{\text{exp}}$ is the $j$th differential cross section point in the $i$th data set,
with electron energy $E_{i,j}$ and four-momentum transfer $Q_{i,j}^2$,
$\sigma(E_{i,j},Q_{i,j}^2)$ is the corresponding differential cross section computed with Eq.~(\ref{cross_section_1PE}).
The uncertainty $\Delta \sigma_{i,j}$ of $\sigma_{i,j}^{\text{exp}}$ includes
the statistical and uncorrelated systematic uncertainties added in quadrature.

We perform a {\emph{simultaneous}} fit of the polarization and cross section data by minimizing
the sum of the least-square functions, Eqs.~(\ref{chi2rat}) and (\ref{chi2_cross_only}):
\begin{equation}
\label{chi2_ratio_cross}
\chi^2 = \chi^2_{\text{rat}} + \chi^2_{\text{cs}}
\,.
\end{equation}

For the fit~I the range of $Q^2$ taken into account is
\begin{equation}
0.1\, \text{GeV}^2 \leq Q^2 \leq 6 \, \text{GeV}^2
\,,
\label{Q2}
\end{equation}
which corresponds to the interval of $Q^2$ values where polarization transfer data are available.

For the  fit~II we extend the range of $Q^2$ down to $Q^2\simeq 0$:
\begin{equation}
0  \leq Q^2 \leq 6 \, \text{GeV}^2
\,.
\label{Q2_second}
\end{equation}
Notice that a precise knowledge of the form factors in the low $Q^2$ region is of special interest for neutrino-nucleon (-nucleus) scattering processes.
In both cases the upper limit, $Q^2=6$~GeV$^2$, is determined by the polarization data. We do not
consider higher $Q^2$ points.

As already stressed in the literature~\cite{Arrington:2007ux}, the inclusion of the most precise data of Ref.~\cite{Qattan:2004ht} definitely indicates the need of  corrections to formula (\ref{cross_section_1PE}). Beyond the  classical radiative corrections \cite{Mo:1968cg}, to get agreement with the polarization data, one needs to consider also the two photon exchange (TPE) corrections, which
can be written  as an additive term  to the reduced cross section:
\begin{equation}
\sigma_R \to \sigma_R + \delta_{TPE}.
\end{equation}

The calculation of $\delta_{TPE}$ is difficult and model dependent: however, one can use general properties to derive a phenomenological expression of the TPE term. The scattering amplitude for electron-nucleon interaction must  satisfy  general symmetry properties, such as crossing symmetry and $C$-invariance~\cite{Rekalo:2003xa}. They can be used to constrain $\delta_{TPE}$.

Following Ref.~\cite{Chen:2007ac} we adopt a TPE correction given by a function $F(Q^2,y)$
\begin{equation}
\sigma_R \to \sigma_R + \tau F(Q^2,y)
\end{equation}
where
\begin{equation}
y = \sqrt{\displaystyle \frac{1-\epsilon}{1+\epsilon} }\,.
\end{equation}
The function $F(Q^2,y)$ must satisfy the relation $F(Q^2,y)= -F(Q^2,-y)$. The analytical properties of  $F(Q^2, y)$ allow one to express this function as a series of odd powers of $y$. Chen \textit{et al.} \cite{Chen:2007ac} truncated the expansion to the second term:
\begin{equation}
F(Q^2, y) = \alpha G_D^2(Q^2) y + \beta G_D^2(Q^2) y^3,
\end{equation}
$\alpha$ and $\beta$ being fit parameter and $G_D(Q^2)$ the usual dipole form factor:
\begin{equation}
G_{D}(Q^2) = \left( 1 + \frac{Q^2}{M_V^2}\right)^{-2}
\,,
\qquad
\text{with}
\qquad
M_V^2 = 0.71\, \mathrm{GeV}^2
\,.
\label{dipol}
\end{equation}
\begin{figure}[t]
\centering{
\includegraphics[width=0.8\textwidth]{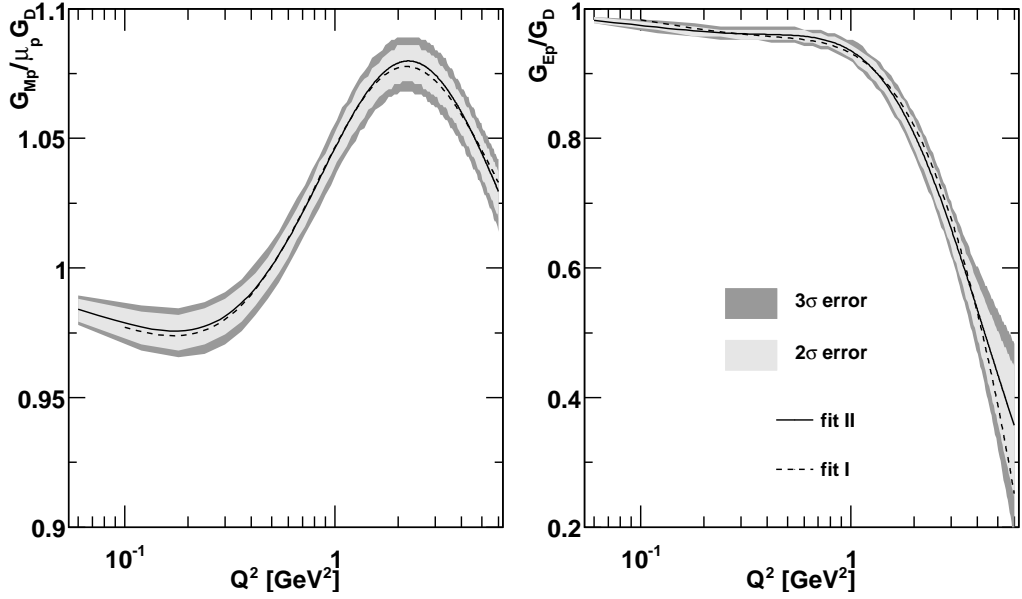}
\caption{ Electric and magnetic proton form factors. The solid lines
denote the  best fit~II. The shadowed areas represent the  $2\sigma$
(bright) and $3\sigma$ C.L. (dark) allowed regions. Our previous
fit~I is shown with the dashed lines.
\label{gmp_and_gep_comparison_r_and_k}}}
\end{figure}

We consider both types of fit; for  the fit~I we obtained:
\begin{equation}
\chi^2_{\text{min}} / \text{NDF} = 375.97 / 392, \quad \mathrm{GoF}=
71\% \,, \label{chi2min-glo_1}
\end{equation}
with the following values for the best fit parameters:
\begin{equation}
a_{p,1}^M = 1.53 \pm 0.01
\,,\;
b_{p,1}^M= 12.87 \pm 0.07
\,,\;
b_{p,2}^M = 29.16 \pm 0.25
\,,\;
b_{p,3}^M = 41.40 \pm 0.33
\,,\;
c_0 = 1.02 \pm 0.01
\,,\;
c_1= -0.13 \pm 0.01
\,.
\label{par-glo_1}
\end{equation}
The parameters of the TPE correction are:
\begin{equation}
\label{par-glo-1-tpe} \alpha = -0.39 \pm 0.09,\quad \beta = -0.04
\pm 0.09\,.
\end{equation}

Notice that the values of $c_0$ and $c_1$ parameters are very similar to the ones in Eq.~(\ref{ratio-fit}), obtained by fitting the polarization transfer data alone.
\begin{figure}[t!]
\centering{
\includegraphics[width=0.6\textwidth]{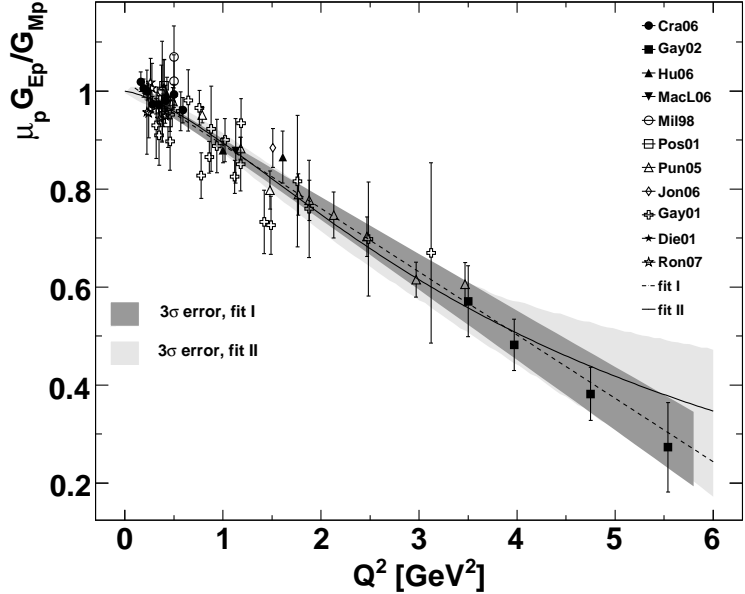}
\caption{ The ratio $\mu_p G_{Ep}/G_{Mp}$ obtained in the
simultaneous fit to polarization measurements and cross section
data. fit~I is shown by the dashed line, the corresponding $3\sigma$
C.L. allowed region being denoted by dark shadowed area. fit~II is
shown by the solid line, the corresponding $3\sigma$ C.L. allowed
region being denoted by bright shadowed area.
\label{ratio-fit-new}}}
\end{figure}

For the fit II the minimization procedure leads to:
\begin{equation}
\chi^2_{\text{min}} / \text{NDF} = 450.95 / 468, \quad \mathrm{GoF}=
71\% \,, \label{chi2min-glo_2}
\end{equation}
with the following values for the fit parameters:
\begin{eqnarray}
b_{p,1}^M&=& 12.31 \pm 0.07
\,,\;
b_{p,2}^M = 25.57 \pm 0.22
\,,\;
b_{p,3}^M = 30.61 \pm 0.27,
a_{p,1}^M = 1.09 \pm 0.01\,,\;
\nonumber\\
b_{p,1}^E&=& 11.12 \pm 0.15
\,,\;
b_{p,2}^E = 15.16 \pm 1.03
\,,\;
b_{p,3}^E = 21.25 \pm 3.27
\,,\;
a_{p,1}^E = -0.19 \pm 0.06.
\label{par-glo_2}
\end{eqnarray}
The parameters of the TPE correction are:
\begin{equation}
\label{par-glo-2-tpe}
\alpha = -0.36 \pm 0.09, \quad \beta = -0.08 \pm 0.09.
\end{equation}

We remark that from both fits we obtained comparable values of
the TPE parameters (see Fig. \ref{fig_contour_tpe_r},  which illustrates the allowed regions in the ($\alpha,\beta$) parameter space with a given confidence level (C.L.)). In both cases
the TPE correction turns out to be negative.
Let us mention that the way we introduce the TPE corrections in our analysis also motivates the choice for the upper $Q^2$ limit: indeed, following the approach of Ref.~\cite{Chen:2007ac}, the magnitude of TPE is \emph{fitted} to the data and, in the elastic cross section, it can be comparable to the magnitude of
$G_{Ep}$. Hence the inclusion of the polarization data (which are less affected by TPE correction) allows a more precise determination of the TPE fit parameters, but restricts the $Q^2$ range to the one of the available polarization data.

It is worth mentioning that by excluding the TPE correction (hence using for the cross section formula (\ref{cross_section_1PE})) both fits worsen, particularly in the goodness of fit. For the fit~I we obtain $\chi^2_{\text{min}} / \text{NDF} = 467.07 / 394$ with $\mathrm{GoF} = 0.6 \%$; similarly for the fit~II we get $\chi^2_{\text{min}} / \text{NDF} = 544.31 / 470$ with $\mathrm{GoF} = 1\%$.
We noticed that this result stems from the presence, in the analysis, of the very accurate JLab data \cite{Qattan:2004ht}, without which GoF would increase to $45\%$ and $47\%$, respectively.

In addition to the above discussed form factors, we also checked a different parameterization, based on a two-poles formula for both the electric and magnetic proton form factors~\cite{Alberico:1995bi}:
\begin{eqnarray}
G_{Ep}(Q^2)&&= \frac{a_1^E}{1+a_2^E Q^2} +\frac{1-a_1^E}{1+a_3^E Q^2}
\label{two-polesE}\\
\frac{G_{Mp}(Q^2)}{\mu_p}&&= \frac{a_1^M}{1+a_2^M Q^2} +\frac{1-a_1^M}{1+a_3^M Q^2}.
\label{two-polesM}
\end{eqnarray}
With respect to Kelly's parameterization this would offer the advantage of having a smaller number of parameters, in addition to the ones of the TPE correction. A new, global fit, can be obtained with
$\chi^2_{\text{min}} / \text{NDF} = 1.10$ but $\mathrm{GoF} = 0.06 \%$, thus indicating that the former parameterization is preferable.

As it has been already mentioned in the introduction, one of our
main tasks is to compute the form factor uncertainties as they can
be extracted from the fit. This goal can be achieved by performing
an accurate error analysis on the various fit parameters.

\begin{figure}[t!]
\centering{
\includegraphics[height=18cm]{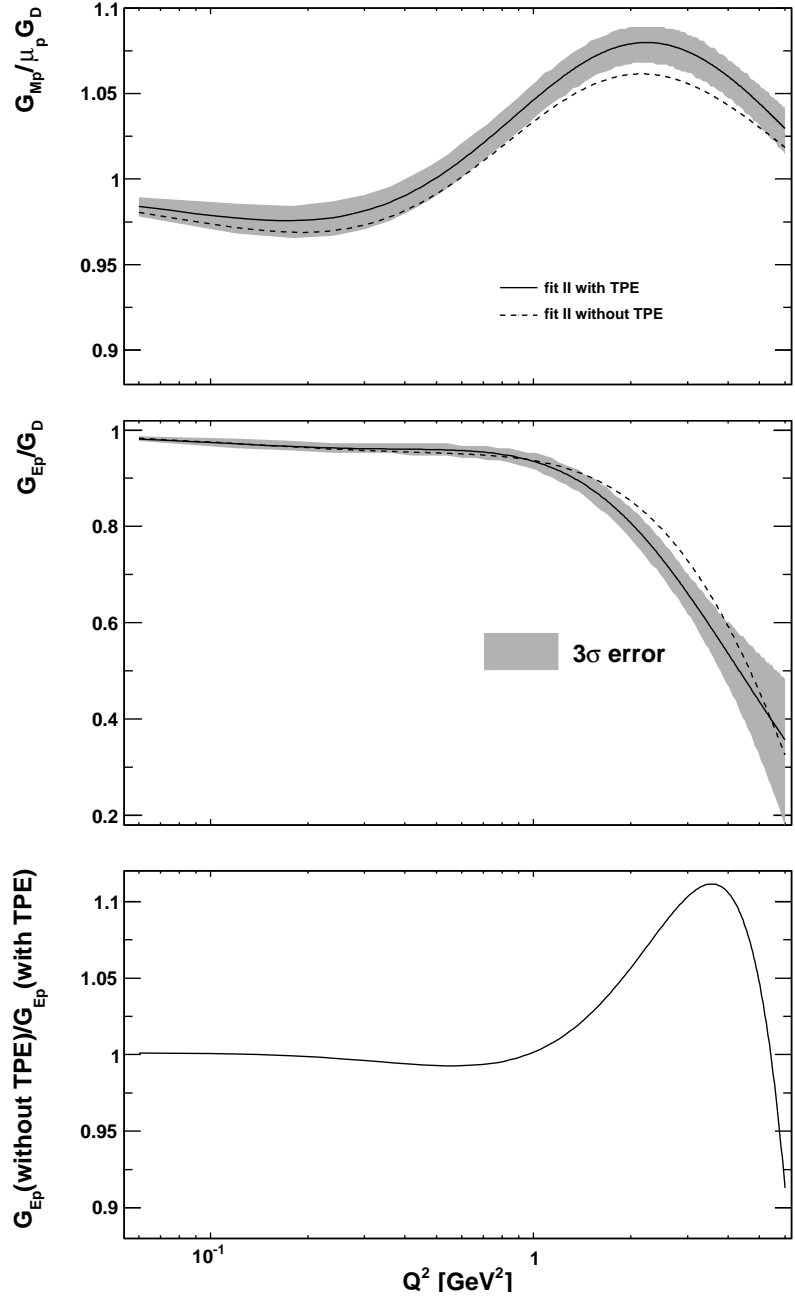}
\caption{ Comparison of fits obtained with and without TPE
correction (solid and dashed lines correspondingly). In the first
and second panels $G_{Mp}/(\mu_p G_D)$ and $G_{Ep}/G_D$ are plotted,
respectively. The shadowed areas denote the $3\sigma$. allowed
regions. In the bottom panel the ratio (\ref{ratio_GE_tpe}) is
shown. Results are obtained for fit~II.
\label{gmp_and_gep_with_and_withot_tpe}}}
\end{figure}

We calculated the correlated uncertainties of the fit parameters and the related uncertainties of the form factors with the standard least-squares method, which is appropriate and widely used for non-linear models\footnote{In this case,
the uncertainties of the parameters and their correlations estimated
from the covariance matrix are quite approximate.
This method was widely used in the past, when computer power was insufficient to perform
more accurate evaluations, as the one presented in this paper.
}
(see Ref.~\cite{Eadie-71,Groom:2000in,Amsler:2008zz}):
the allowed region in the space of $N$ parameters with $\lambda$ confidence level (C.L.)
is delimited by the contour defined by
\begin{equation}
\chi^{2} = \chi^{2}_{\text{min}} + \Delta\chi^{2}(N,\lambda)
\,,
\label{contour}
\end{equation}
where $\Delta\chi^{2}(N,\lambda)$ is the value for which a $\chi^2$ variable with $N$ degrees of freedom has a cumulative probability $\lambda$.
We consider $2\sigma$ (95.45\% C.L.) and $3\sigma$ (99.73\% C.L.) uncertainties.

Since we have 8 parameters in fit I and 10 parameters in fit II,
the exploration of the parameter space in order to find the contours defined by Eq.~(\ref{contour})
cannot be done with the simplest grid method.
Therefore,
we used a Monte Carlo Markov Chain generator of random points,
which allows to find the allowed parameter regions with good accuracy in a few hours of CPU time of a normal PC.

It is interesting to notice, that the estimated values of the magnetic
form factor parameters, $a_{p,1}^M$, $b_{p,1}^M$, $b_{p,2}^M$ and
$b_{p,3}^M$ are strongly correlated. In particular, the estimates
of $a_1$ and $b_3$ are almost linearly dependent. These parameters
determine the asymptotic behavior of $G_{Mp}(Q^2)$, which turns out
to be:
\begin{equation}
\lim_{Q^2\to \infty } Q^4 G_{Mp}(Q^2)
=
(4M^2)^2 \, \frac{a_{p,1}^M}{b_{p,3}^M}
\simeq
( 0.68^{\,+0.01}_{\,-0.01} \, \mathrm{GeV}^2 )^2
\,,
\end{equation}
in fair agreement with the one given by the usual dipole form factor
(\ref{dipol}) (the above uncertainties are at $3\sigma$).

In Fig.~\ref{fig_contour_tpe_r} (left panel) the error contours for
$\alpha$ and $\beta$ parameters are  shown. Let us notice that
solutions with $\beta$ positive but very small are possible, but in
this case  $\alpha$ should be negative and large in magnitude.
Therefore the TPE correction are always negative.

In Fig.~\ref{gmp_and_gep_with_ratio} we show our best fits  (fit~I)
for the magnetic and electric proton form factors with their
uncertainties. We compare with Kelly's fit \cite{Kelly:2004hm}.
One can see that Kelly's fits of the magnetic proton form
factor lies within our $3\sigma$ C.L. region in almost the whole $Q^2$ range
under consideration. For the electric form factor the fits differ by
more than $3\sigma$ in a relatively wide range of $Q^2$.

A similar error analysis is performed for fit~II. Here, the number
of form factor parameters is larger (4 parameters for each form
factor). Similarly as above we show  contour plot for the TPE
correction parameters (Fig.~\ref{fig_contour_tpe_r}, right panel).

Analogously to the case of fit~I the estimates of the parameters
$a_{p,1}^M$ and $b_{p,3}^M$ are linearly dependent and their ratio
is similar:
\begin{equation}
\lim_{Q^2\to \infty } Q^4 G_{Mp}(Q^2)
=
(4M^2)^2 \, \frac{a_{p,1}^M}{b_{p,3}^M}
\simeq
( 0.66^{\,+0.02}_{\,-0.02} \, \mathrm{GeV}^2 )^2.
\,
\end{equation}
{
One could also derive from the form factor parameters the charge and magnetic
root-mean-square radii for the proton, as given by the slope of the electric
and magnetic form factors at $Q^2=0$. They turn out to be:
$\sqrt{<r_{Ep}^2>} = 0.87\pm 0.01\, \mathrm{fm}, \quad \sqrt{<r_{Mp}^2>} = 0.86\pm 0.01, \, \mathrm{fm}$.
These results are comparable with previous analysis in the literature, but
slightly lower than the most recent and advanced estimates of Ref.~\cite{Kelly:2002if,Sick:2003gm}; indeed the latter take into account Coulomb distortion, which is relevant at low  $Q^2$. For example Ref.~\cite{Sick:2003gm} provides $\sqrt{<r_{Ep}^2>} = 0.895\pm 0.018\, \mathrm{fm}$. The present fit of the form factors is carried out in plane-wave approximation and low-$Q^2$ properties like charge
and magnetic radii are not properly reproduced without accounting for radiative corrections to the Rosenbluth cross sections.}

In Fig.~\ref{gmp_and_gep_comparison_r_and_k} we present $G_{Mp}$ and
$G_{Ep}$ obtained in fit~II, with the $2\sigma$ and $3\sigma$ C.L.
error bands (represented  by shadowed areas). Here the results from
fit~I are also plotted. Both fits lead to very similar magnetic form
factors. On the contrary, there is a visible difference between the
corresponding electric form factors: the $G_{Ep}$ obtained in fit~I
decreases faster then the one obtained in fit~II.

Fig.~\ref{ratio-fit-new} shows the ratio $\mu_p G_{Ep}/G_{Mp}$
obtained with the fits I and II. The  linear ratio fitted only to
the polarization data  is no longer shown, since it is very similar
to the one obtained with fit~I. The ratio uncertainties  are larger
for the fit~II than for fit~I due to the fact that the
parameterization in fit~II contains a larger number of degrees of
freedom.

Finally, given for granted that they are necessary, it is interesting to understand
which is the \emph{quantitative} impact of the
TPE correction: they are expected to be relevant especially for the
electric form factor. For this purpose we compare in
Fig.~\ref{gmp_and_gep_with_and_withot_tpe} the proton form factors
obtained with and without TPE correction  -- only fit~II is
considered. The magnetic proton form factor obtained without TPE is
systematically shifted down by about 1.5\%, and above
$Q^2 \simeq 1$ GeV$^2$ it lies outside the $3\sigma$ C.L. region of the
form factor obtained by including the TPE correction. The analogous
effect on the electric form factor is shown in the middle panel of
the same figure and appears to be less uniform than for the magnetic
form factor: this can be better appreciated from the bottom panel of
Fig.~\ref{gmp_and_gep_with_and_withot_tpe}, where the ratio
\begin{equation}
\label{ratio_GE_tpe}
G_{Ep}(\mathrm{without}\, \mathrm{TPE})/G_{Ep}(\mathrm{with}\, \mathrm{TPE})
\end{equation}
is plotted. One can see that the TPE correction substantially alters
the $Q^2$ dependence of the electric form factor, in particular,
for $Q^2>2$~GeV$^2$, with an effect which grows up to the order of
10\%. In any case the impact of the TPE correction turns out to be
non-negligible for both form factors.

\begin{figure}[t!]
\centering{
\includegraphics[width=0.6\textwidth]{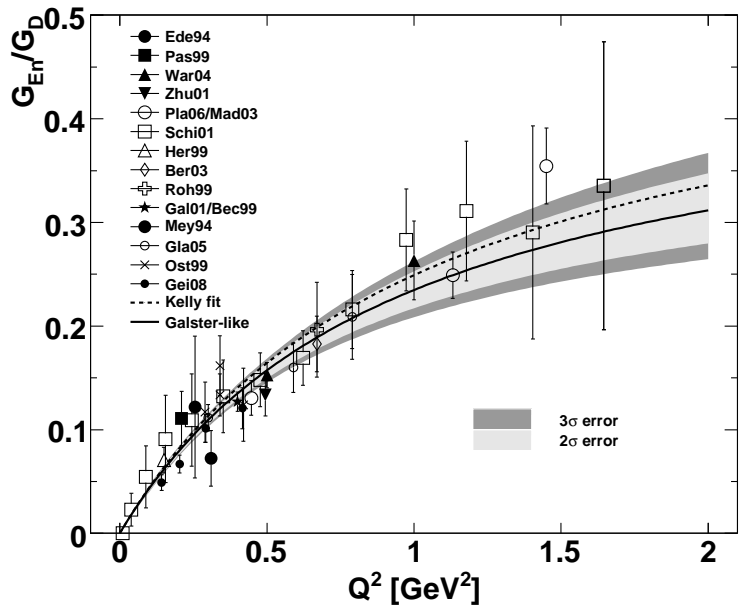}
\caption{ Fit of the electric neutron form form factor (solid line).
The Galster-like parameterization (\ref{GEn_par}) is considered. The
shadowed areas denote the $2\sigma$ (bright) and $3\sigma$ C.L.
(dark) allowed regions. Kelly's fit \cite{Kelly:2004hm} is shown
with the dotted line. \label{fig_gen_galster} } }
\end{figure}

\section{Neutron form factors}

The measurement of the neutron form factors is much more difficult
than that of the proton form factors, since  a target  of free
neutrons  does not exist. The neutron form factors are extracted
from measurements of electron-nucleus scattering, usually
electron-deuteron or electron-helium scattering. Therefore, the data
analysis is affected by uncertainties stemming from the nuclear
theoretical model assumed to describe the target nucleus. Since
these models have consistently improved with time, in our analysis
we consider only relatively recent data. At variance with the proton
case, we take from the literature directly the published values of
neutron form factors ``data'' and apply our fitting procedure to
them.

\subsection{Electric neutron form factor}

\begin{figure}[t!]
\centering{
\includegraphics[width=0.6\textwidth]{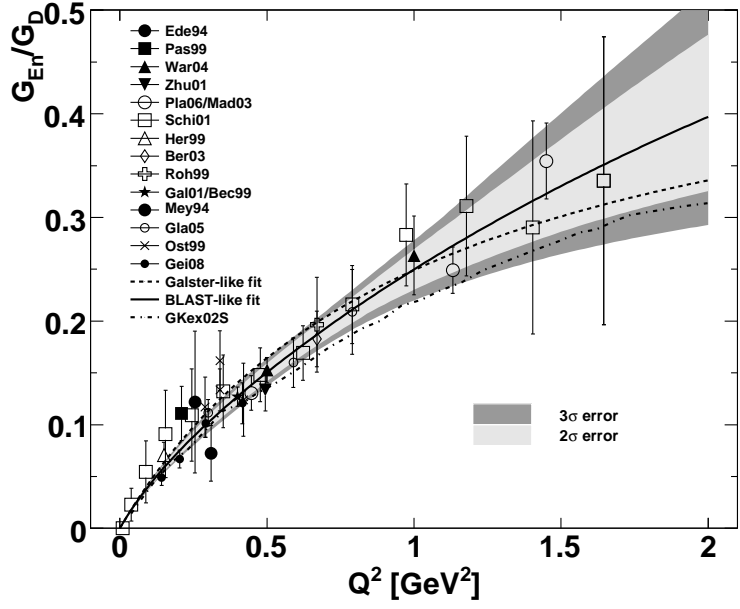}
\caption{ Fit of the electric neutron form form factor (solid line).
The BLAST-like parameterization (\ref{gen_par_blast}) is shown. The
shadowed areas represent the $2\sigma$ (bright) and $3\sigma$ C.L.
(dark) allowed regions. The dashed line denotes the Galster-like
parameterization. The Lomon result (GKex02S) \cite{Lomon:2006xb} is
denoted by the dash-dotted line. \label{fig_gen_blast} } }
\end{figure}

For the electric neutron form factor we adopt the Galster-like parameterization
\begin{equation}
\label{GEn_par}
G_{En}(Q^2) = \frac{A \tau}{1 + B \tau} \, G_{D}(Q^2)
\,,
\end{equation}
with the dipole form factor of Eq.~(\ref{dipol}).
\begin{figure}[t!]
\centering{
\includegraphics[width=0.6\textwidth]{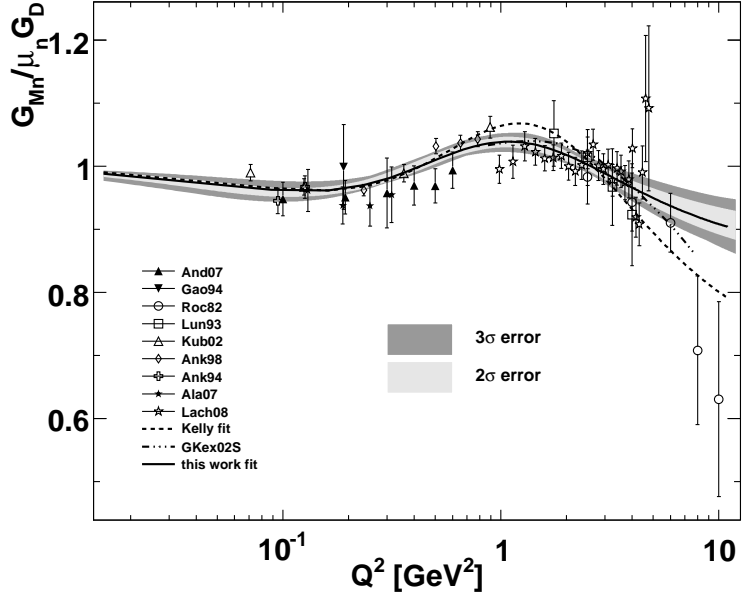}
\caption{Fit of magnetic neutron form factor (solid line),
normalized to the dipole form factor. The shadowed areas denote
$2\sigma$ (bright) and $3\sigma$ C.L. (dark) allowed regions.
Kelly's fit \cite{Kelly:2004hm} is shown with dotted line.
\label{fig_gmn}} }
\end{figure}

We consider the electric neutron form factors ``data'' which
have been published in several papers. Some of the data have been
obtained in asymmetry and recoil polarization measurements
\cite{Bermuth:2003qh,Golak:2000nt,Becker:1999tw,Meyerhoff:1994ev,Rohe:1999sh,Zhu:2001md,Eden:1994ji,Herberg:1999ud,Passchier:1999cj,Warren:2003ma,Glazier:2004ny,Ostrick:1999xa,Plaster:2005cx,Madey:2003av}.
We consider also the reanalyzed electron-deuteron data
\cite{Schiavilla:2001qe} and the newest BLAST measurements~\cite{BLAST:2008ha}.
Additionally, in order to have a
proper slope of the electric form factor in the limit $Q^2\to 0$, we
impose to our fit the additional constrain \cite{Kopecky:1995zz}:
\begin{equation}
\left<r_{En}^2\right> = -0.1148 \pm 0.0035~\mathrm{fm}^2.
\end{equation}

We considered a least-squares function similar to the one in Eq.~(\ref{chi2rat}),
with the experimental statistical and systematical uncertainties added in quadrature.
With the values:
\begin{equation}
A =1.68 \pm 0.05
\,,
\qquad
B = 3.63 \pm 0.39
\label{par-GEn}
\,,
\end{equation}
we obtained
\begin{equation}
\chi^2_\mathrm{min} / \mathrm{NDF} = 25.82/37 \,,
\label{chi2min-GEn}
\end{equation}
and the goodness of the fit turned out to be excellent: 91\%.

In Fig.~\ref{fig_gen_galster} we plot the best-fit value of $G_{En}$
as a function of $Q^2$ together with the 2$\sigma$ and 3$\sigma$
C.L. allowed regions. We plot also Kelly's fit~\cite{Kelly:2004hm}.

As an alternative to the most commonly used Galster-like
parameterization, we considered a neutron electric form factor given
by the sum of two dipole form factors:
\begin{equation}
\label{gen_par_blast} G_{En}(Q^2) = \displaystyle \frac{a}{\left( 1
+ b_1 Q^2 \right)^2} - \frac{a}{\left( 1 + b_2 Q^2 \right)^2}.
\end{equation}
This parameterization is similar to the one considered in the latest
BLAST data analysis \cite{BLAST:2008ha} and for this reason we will
call it BLAST-like parameterization. The fitting procedure for the
above parameterization leads to:
\begin{equation}
\label{chi2min-gen-blast} \chi^2/NDF = 17.95/36, \quad \mathrm{GoF}
= 99\%
\end{equation}
with the parameters:
\begin{equation}
\label{par-gen-blast} a = -0.10 \pm 0.02, \quad b_1 = 2.83 \pm 0.37
, \quad b_2 = 0.43 \pm 0.11.
\end{equation}

In Fig.~\ref{fig_gen_blast} the BLAST-like parameterization is
compared to the data: the Galster-like parameterization and one of
the recent Lomon parameterization~\cite{Lomon2001,Lomon:2006xb}
(GKex02S) are also shown. Notice that the parameterization
(\ref{par-gen-blast}) raises faster with $Q^2$ than the Galster-like
and the Lomon one, but the latter remain both within the BLAST
parameterization uncertainties.

\subsection{Magnetic neutron form factor}

\begin{figure}[t!]
\centering{
\includegraphics[width=0.8\textwidth]{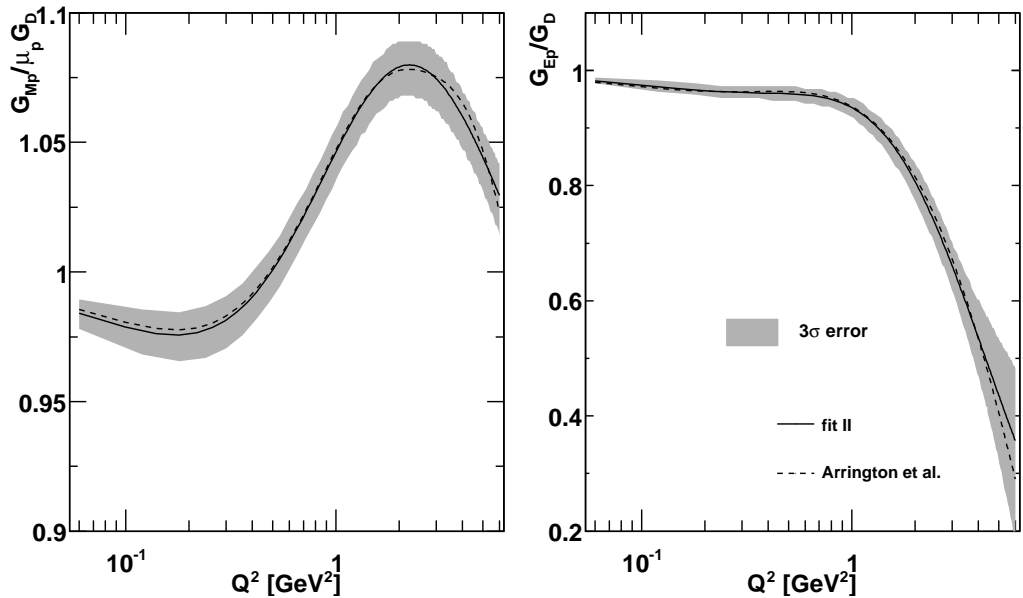}
\caption{ Magnetic and electric proton form factors (normalized to
the dipole form factor). The comparison between fit~I (solid lines)
and Arrington \textit{et al.}~\cite{Arrington:2007ux} (dashed lines)
is shown. The shadowed areas denote the $3\sigma$ C.L. allowed
region. \label{fig_comparisons_1}} }
\end{figure}

For the neutron magnetic form factor we adopted again the simplest form of Kelly's parameterizations, with $n=1$:
\begin{equation}
\label{Kelly_Form_Factor_neutron}
\frac{ G_{Mn}(Q^2) }{ \mu_n }
=
\frac{\displaystyle 1 +  a_{n,1}^{M}\tau }{\displaystyle 1 + b_{n,1}^{M}\tau + b_{n,2}^{M}\tau^2 + b_{n,3}^{M}\tau^3}
\,.
\end{equation}
We considered 11 data sets, obtained from asymmetry measurements
\cite{Anderson:2006jp}-\cite{Gao:1994ud},
\cite{BLASTAlarcon:2007zza} and cross section measurements in
electron-deuterium scattering
\cite{Anklin:1994ae}-\cite{Lachniet:2008qf}, where Ref.
\cite{Lachniet:2008qf} contains the latest JLab measurements. The
fit to all these data sets leads, however, to a minimum
$\chi^2/\text{NDF}=2.05$, not quite satisfactory. According to a
remark of Kelly~\cite{Kelly:2002if}, the data from
\cite{Markowitz:1993hx} and \cite{Bruins:1995ns} were extracted
using the same associated-particle technique for the neutron
efficiency, a technique which appears to be in contradiction with
the method used in other experiments.  Therefore we omitted these
two data sets. From our final analysis. After excluding the two
above mentioned data sets, we obtained a fit over $N=56$ points with
\begin{equation}
\label{chi2min-GMn} \chi^2_{min}/\text{NDF}= 52.79/52 = 1.01,
 \end{equation}
and $\mathrm{GoF} = 44\%$. The corresponding values for the parameters in the neutron magnetic form factor are:
\begin{eqnarray}
\label{GMn-par}
b_{n,1}^M &=& 21.30  \pm 4.56,  \quad
b_{n,2}^M  = 77     \pm 31 \quad
b_{n,3}^M  = 238    \pm 105 ,\quad
a_{n,1}^M  = 8.28 \pm 3.89.
\end{eqnarray}
\begin{figure}[t!]
\centering{
\includegraphics[width=\textwidth]{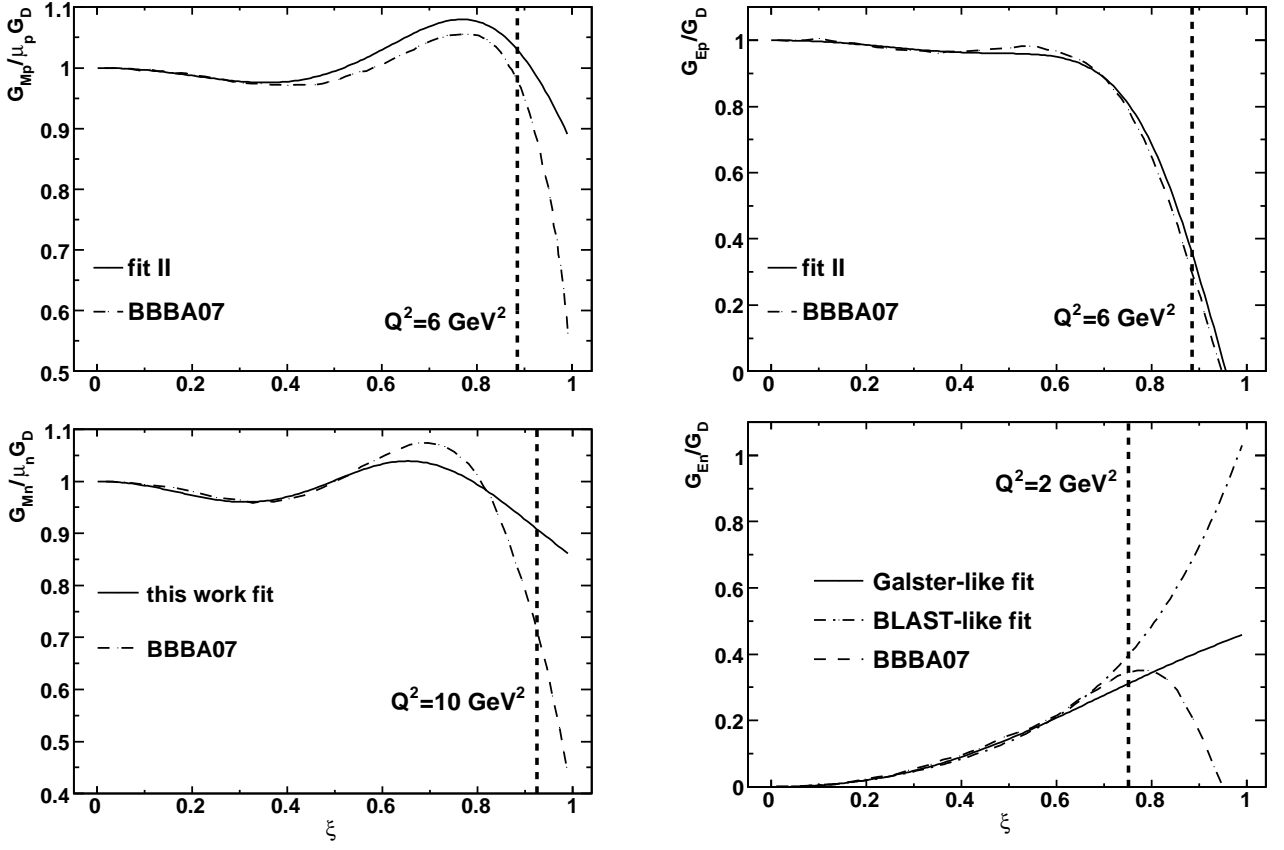}
\caption{Proton and neutron electric and magnetic form factors (normalized to the dipole form factor). The fits  of electric and magnetic form factors (fit II) as well as the fit of neutron magnetic form factor are denoted by solid lines. BBBA07 \cite{Bodek:2007ym} global fits are
denoted by dashed lines. In the case of the electric neutron form factor, both parameterization, Galster and BLAST -like, are shown by the solid and dash-dotted lines, respectively.  The form factors are plotted against the Nachtman variable. The dashed vertical bar represents the upper bound of the $Q^2$ range were our fits were performed.
\label{fig_comparison_with_bodek}}
}
\end{figure}

We performed the error analysis over the four parameters above. Even
if the fit of $G_{Mn}$ was done on a slightly different basis than
the one of the proton, yet we observe strong correlations between
estimated values of the parameters, in full analogy with our
findings for the proton form factor. In particular, the estimated
values of parameters ($a_1$ and $b_3$) which determine the
asymptotic behavior of the form factor at large $Q^2$ are almost
linearly dependent:
\begin{equation}
\label{GMn_limit} \lim_{Q^2\to \infty } Q^4 G_{Mn}(Q^2) = (4M^2)^2
\, \frac{a_{n,1}^M}{b_{n,3}^M} \simeq (0.66^{\,+0.01}_{\,-0.01} \,
\mathrm{GeV}^2)^2
\end{equation}
This value is very similar to the one obtained for the
proton. Notice, however, that without the newest JLab
data, instead of the value (\ref{GMn_limit}) we would get
$(0.58^{\,+0.04}_{\,-0.05}\, \mathrm{GeV}^2)^2 $.
In Fig.~\ref{fig_gmn} our final fit of $G_{Mn}$ is different
from Kelly's result~\cite{Kelly:2004hm} since it contains the
newest JLab measurements.

\begin{figure}[t!]
\centering{
\includegraphics[width=16cm, height=7cm]{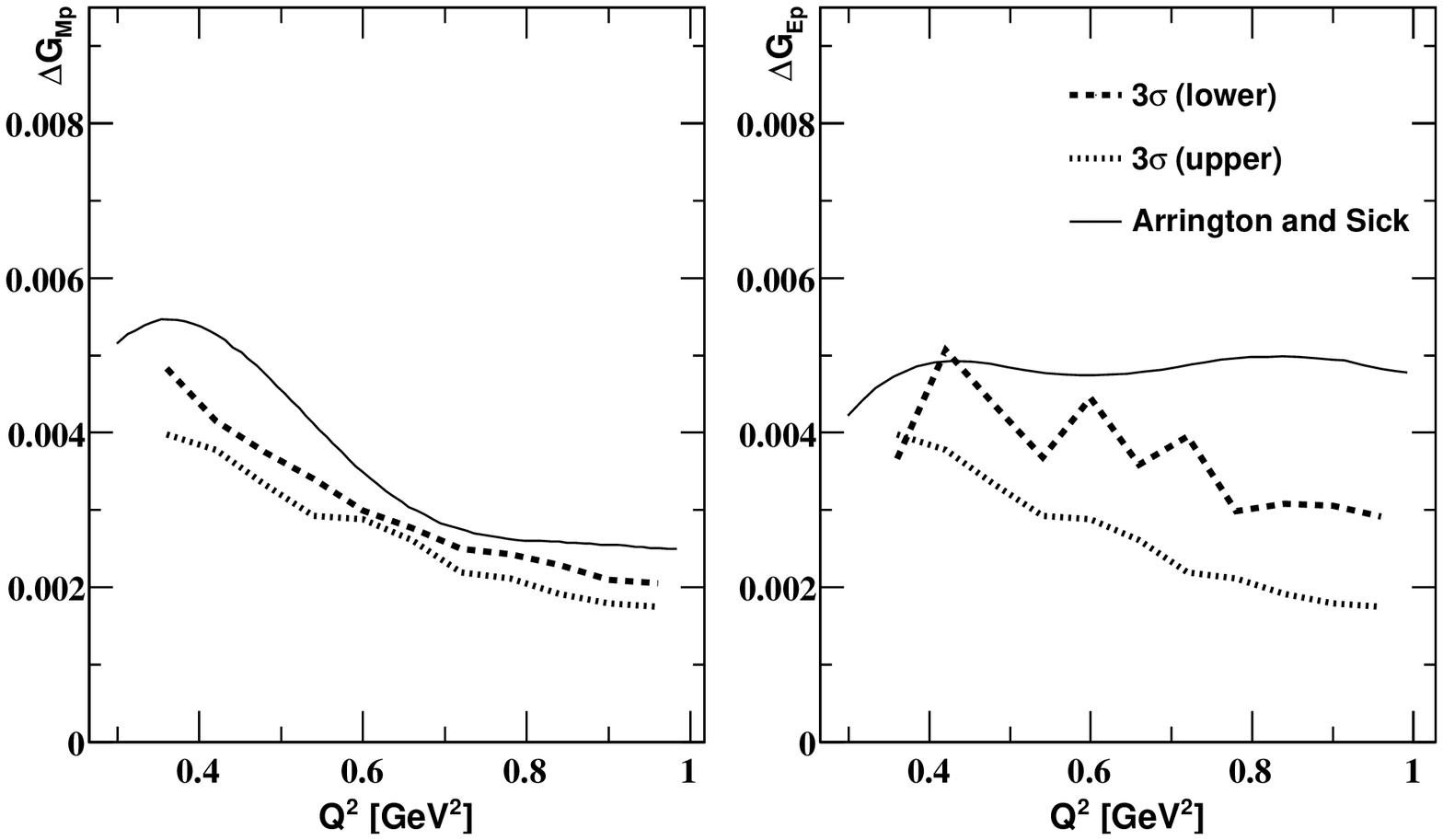}
\caption{ The uncertainties of electric and magnetic proton form
factors. Our computations (fit II) are denoted by dotted (upper
uncertainties) and dashed lines (lower uncertainties) for the
$3\sigma$ C.L. errors. The solid lines correspond to the results
obtained in
Ref.~\cite{Arrington:2006hm}.\label{fig_comparison_of_errors} } }
\end{figure}

\section{Discussion and conclusions}

In this section we start by presenting further comparisons of the
form factors resulting from our fits with the ones of previous data
analyses.

In Ref.~\cite{Arrington:2007ux} the first systematical
global analyses of the cross section and polarization transfer data
on the proton  with the inclusion of the TPE correction was
performed. That fit is valid up to $Q^2=30$~GeV$^2$ for the magnetic
form factor and up to 6~GeV$^2$ for the electric form factor. In
Fig.~\ref{fig_comparisons_1} we display together our global fits and
those of Ref.~\cite{Arrington:2007ux}: it clearly appears that, even
thought different approaches for the TPE correction were employed,
the global fits are very similar.

We also compare our fits with the recent one of Bodek \textit{et
al.}~\cite{Bodek:2007ym} (BBBA07). This global fit is tailored to
accurately describe the form factors at low $Q^2$ as well as in the
intermediate region of $Q^2$. These authors used the Kelly's
parameterization with four parameters but each form factor was
multiplied by some Legendre polynomial, which depends on several
additional parameters, constrained to reproduce the low $Q^2$
behavior obtained in Ref.~\cite{Arrington:2006hm}.

The authors of Ref.~\cite{Bodek:2007ym} plotted the form factors
against the so-called Nachtman variable, {which for the elastic
scattering is defined as} $ \xi = 2/(1 + \sqrt{ 1 +1/\tau} )$.
Therefore in order to make the comparison with their results we
express our form factors in terms of the $\xi$ variable. Plots are
shown in Fig.~\ref{fig_comparison_with_bodek}. In the region of
$\xi$ corresponding to the range of validity of our fits the
predictions of the two parameterizations are very similar, however,
our magnetic proton form factor is systematically higher (by several
percent) then the one given by the BBBA07 parameterization. The
difference is given by TPE correction which we considered in our
fitting procedure (see Fig. \ref{gmp_and_gep_with_and_withot_tpe}),
while authors of Ref. \cite{Bodek:2007ym} did not discuss this
effect. For higher $\xi$ values one can notice sizeable deviations
for our form factors. However, we notice, that the $\xi$ variable
compresses in a very short range the large $Q^2$ region.

Finally we compare our estimates of uncertainties with those
obtained by Arrington and Sick~\cite{Arrington:2006hm}: these
authors did a serious attempt to compute the uncertainties of the
nucleon form factors, which is a crucial information in the study of
parity violating $ep$ scattering. Their method to compute errors is
explained in Ref.~\cite{Arrington:2006hm} and differs from our,
in particular in the treatment of the systematic uncertainties.
In particular we usually obtain asymmetric uncertainties around the
best fit value: hence in Fig. \ref{fig_comparison_of_errors} we
compare the errors of Arrington and Sick on the electric and
magnetic proton form factors with our lower and upper bounds for the
$3\sigma$ confidence level errors. For the magnetic proton form
factor, within the $3\sigma$ C.L. our results are consistent with
the ones of Ref.~\cite{Arrington:2006hm}. For the electric proton
form factor we notice some deviations between our results and the
ones obtained by Arrington and Sick.

In conclusion we have presented two fits of the proton and neutron
electromagnetic form factors, using the best available data. The
$ep$ elastic cross sections were reproduced by including a simple
but realistic parameterization of the two photon exchange correction.
Alternative parameterizations with fewer parameters than the
one employed here do not allow to obtain equally good fits.
We show that the impact of the TPE correction on the magnetic and
electric proton form factors  is larger than the $3\sigma$ uncertainty
of the fits (in a wide range of $Q^2$).  In fit~I we constrained the
electric proton form factor by the ratio $\mu_p G_{Ep}/G_{Mp}$
extracted from recoil polarization and asymmetry data. fit~II
employs Kelly's parameterization with four parameters both for the
electric and magnetic proton form factors. This fit is obtained with
two additional parameters  with respect to fit~I, however we believe
that it is more reliable than the former, particularly in the low
$Q^2$ region. We also performed a careful analysis of the
uncertainties resulting on the parameters of the fit and hence on
the form factors. It is worth stressing that only a few papers,
among the many devoted to the nucleon electromagnetic form factors,
include the TPE correction in the analysis. As a final remark we
remind the reader that even small uncertainties in the magnetic form
factors of the proton and neutron turn out to be important for a
correct analysis of the neutrino-nucleon cross sections.

The numerical results of our fits are available in the web site \cite{website_results}.

\section*{Acknowledgements}
The authors acknowledge fruitful discussions with E. L. Lomon and T. W. Donnelly.
K.~M.~G. thanks Ron Gilman for his remarks on the
previous version of the manuscript, which helped to improve the
present one.

S.~M.~B. has been supported by funds of the Munich Cluster of
Excellence (Origin and Structure of the Universe), the DFG
(Transregio 27: Neutrinos and Beyond) and by INFN Sezione di Torino.

K.~M.~G. was supported by WWS project founds.

\end{document}